\documentstyle[prl,aps]{revtex}
\begin{document}
\title{\LARGE \bf THE REAL TEST OF ``TRIVIALITY''  \\
\vspace*{3mm}
            ON THE LATTICE }

\author{ A. Agodi, G. Andronico and M. Consoli}

\address{ Dipartimento di Fisica - Universita di Catania; \\
 Istituto Nazionale di Fisica Nucleare - Sezione di Catania;\\
Corso Italia, 57 - I 95129 Catania - Italy}
\newcommand{\beq}{\begin{equation}}
\newcommand{\eeq}{\end{equation}}

\maketitle

\begin{abstract}
The generally accepted ``triviality'' of $\lambda\Phi^4$ theories does not
forbid Spontaneous Symmetry Breaking but implies a trivially free shifted field
which becomes effectively governed by a quadratic hamiltonian.
As a consequence, one expects the one-loop potential to be
exact . We present a lattice computation of
the effective potential for massless $\lambda\Phi^4$ theory which nicely
confirms the expectations based on ``triviality''.
Our results imply that
 the magnitude of the Higgs boson mass, beyond perturbation theory,
 does not represent a measure of
the observable interactions in the scalar sector of the standard model.
\end{abstract}
\vskip 15 pt
\widetext
 Recently \cite{consteve,zeit} it has been enphasized that the generally
accepted
``triviality''\cite{froh} of $(\lambda\Phi^4)_4$ is not in contradiction
with Spontaneous Symmetry Breaking (SSB). Indeed, ``triviality'' means that
the renormalized Green's functions of the continuum theory can be expressed in
terms of the first two moments of a gaussian distribution. This statement,
by its very nature, does not forbid SSB but implies a trivially free shifted
field $h(x)$ (where $\Phi(x)=\langle \Phi \rangle +h(x)$ )
which becomes effectively controlled by a quadratic hamiltonian.

\par Therefore, it is not surprising that
precisely this result has been obtained in
\cite{cast,con,iban,new}
 by analyzing the effective potential of massless $\lambda\Phi^4$
theories where
SSB \cite{cian,return} is discovered
in those
approximations to the effective potential (one-loop and gaussian)
where the ``Higgs'' field $h(x)$
 is free {\it by definition}. In fact, at one-loop the effects
from the shifted field self- interaction are consistently neglected whereas in
 the gaussian approximation they are fully reabsorbed into the Higgs mass and
the vacuum expectation value by obtaining a normal-ordered quadratic
Hamiltonian
which is equivalent to the Hartree-Fock-Bogolubov approximation.
\par ``Triviality'' implies the inconsistency of the usual
perturbative approach to massless $\lambda\Phi^4$, which is
 based on the attempt to generate a cutoff-independent and
{\it non vanishing } renormalized coupling at non-zero external momenta.
 One should beware of spurious contradictions which arise
by using quasi-perturbative
approximation methods to the effective potential which
are inherently incompatible
with the ``generalized free field'' nature of the field $h(x)$.
Thus,
perturbation theory, the loop expansion (beyond one-loop), and
leading-log re-summation are all wholly misleading because they insist
upon having a finite connected 4-point function at non-zero external
momenta.
The ``triviality'' of $(\lambda\Phi^4)_4$ theory implies
that the bare interaction hamiltonian for the shifted field
(the terms proportional to $h^3$ and $h^4$ ) does not produce any observable
interactions. One can neglect it completely or treat it exactly. However,
it is disastrous to take it into account in a
perturbative or quasi-perturbative manner.
\par To provide further evidence on the validity of our picture
 we have undertaken the lattice computation of the effective
potential along the same lines of Huang et al. \cite{huang}.
In terms of the bare vacuum field $\phi_B= \langle \Phi_B \rangle$ and
 of the bare coupling $\lambda_B$ we obtain the well-known result
\cite{CW}
($\omega^2(\phi_B)={{\lambda_B\phi^2_B}\over{2}}$
and $\Lambda$ is the euclidean ultraviolet cutoff):
\beq
 V^{{\rm 1-loop}}(\phi_B)  =  \frac{\lambda_B}{4!} \phi^4_B +
\frac{\omega^4{\scriptstyle (\phi_B)}}{64\pi^2}
\left( \ln \frac{\omega^2{\scriptstyle (\phi_B)} }{\Lambda^2} -
\frac{1}{2} \right) .
\eeq
This represents the classical potential plus the zero-point energy of the
{\it free} $h$ field. ``Triviality'' implies that there are no $h$-interaction
effects, so that this result should be exact \cite{consteve,zeit}.
 $V^{\rm 1-loop}$
 can be re-expressed in the form
\beq
V^{\rm 1-loop}(\phi_B)={{\pi^2\phi^4_B}\over{Z^2_{\phi} }}
(\ln{{\phi^2_B}\over{v^2_B}}-{{1}\over{2}}) ,
\eeq
where $Z_{\phi}={{16\pi^2}\over{\lambda_B}}$
 from the minimum condition
\beq
 m^2_h =\omega^2{\scriptstyle (v_B)}  =
\frac{1}{2} \lambda_B v^2_B =
\Lambda^2~\exp (-{{32\pi^2}\over{3\lambda_B}}).
\eeq
By introducing the physical vacuum field $\phi^2_R={{\phi^2_B}\over{Z_{\phi}}}$
which reabsorbs into its normalization all zero-momentum interaction effects,
i.e. such that
\beq
   {{d^2V^{\rm 1-loop}}\over{d\phi^2_R}}|_{\phi_R=v_R}=m^2_h
\eeq
we obtain \cite{consteve,con,iban,new}
\beq
V^{\rm 1-loop}(\phi_R)=\pi^2\phi^4_R
(\ln{{\phi^2_R}\over{v^2_R}}-{{1}\over{2}})
\eeq
and $m^2_h=8\pi^2v^2_R$.
The non-perturbative nature of the vacuum field
renormalization ($Z_{\phi}\sim 1/\lambda_B$), first discovered in the
gaussian approximation by Stevenson and Tarrach \cite{return}, should not be
confused with the $h$-field wave function renormalization for which at one
loop, where $h$ is a free field with mass $\omega(\phi_B)$,
 one has trivially $Z_h=1$. The structure
$Z_{\phi}\neq Z_h$, allowed by the Lorentz-invariant nature of the field
decomposition into $p_{\mu}=0$ and $p_{\mu}\neq 0$ components \cite{consteve}
, is more
general than in perturbation theory and is the essential ingredient which
allows SSB to coexist with ``triviality''.
\par Differentiating Eq.(2), we obtain the bare ``source''
\beq
J(\phi_B)={{dV^{\rm 1-loop}(\phi_B)}\over{d\phi_B}}=
{{4\pi^2\phi^3_B}\over{Z^2_{\phi} }}\ln{{\phi^2_B}\over{v^2_B}} ,
\eeq
which can be compared
 with the lattice results for $J=J(\phi_B)$ along the same lines
as Ref. \cite{huang}.
Strictly speaking, just as the effective potential
is the convex envelope of Eq.(2) \cite{convex,rit},
Eq.(6) is valid only for $|\phi_B|>v_B$ and
$J(\phi_B)$ should vanish in the
presence of SSB in the range $-v_B \leq \phi_B \leq v_B$. This means that
 the average bare field
\beq
              \phi_B(J)=\langle\Phi_B\rangle_J~,
\eeq
which,
for any $J\neq 0$,
should satisfy the relation
\beq
             \phi_B(J)=-\phi_B(-J)~,
\eeq
should tend to the limits
\[
  \pm v_B=\lim_{J\to 0^{\pm}}~\phi_B(J).
\]

In our case (we run on a $10^4$ lattice),
Eq.(8) is not well reproduced numerically at low values of $J$
(for $|J|a^3 \sim 0.01$ or smaller, $a$ denoting the lattice spacing).
As a consequence,
the values of $\phi_B$ (as well as of the higher-order Green's functions)
extracted from the direct computation in the
broken case at $J\sim 0$ are not reliable.
Thus, we have to restrict to a ``safe'' region
of $J$-values (in our case $|J|a^3 \geq 0.05$)
 and extrapolate the values of $v_B$ and
$Z_{\phi}$ from a fit to the data by using Eq.(6) once we have identified on
the
lattice the ``massless'' regime, i.e. corresponding to a renormalized theory
with no intrinsic scale in its symmetric phase $\langle \Phi_B\rangle$=0.
Usually, this would require to find the value of the bare mass-squared $r_o$
in the euclidean lattice action
\beq
 a^4 \sum _x~[ {{ \sum^4_{\mu=1}( \Phi_B(x+ae_{\mu})- \Phi_B(x))^2}
\over{2a^2}}
 +
{{1}\over{2}} r_o \Phi_B^2(x) + \frac{\lambda_o}{4} \Phi_B^4(x)]
\eeq
for which
${{d^2V}\over{d\phi^2_B}}|_{\phi_B=0}=0$ \cite{CW}
(in Eq.(9)
we are using the same notations of \cite{huang} and replace in the fourth order
coupling ${{\lambda_B}\over{4!}} \to {{\lambda_o}\over{4}}$ such that
$\lambda_o=1$ implies $\lambda_B=6$). However, the region around $\phi_B=0$
being not directly accessible, we have argued as follows.
 We start with the general expression \cite{zeit}
\beq
J(\phi_B) = \alpha \phi_B^3 \ln(\phi_B^2/v_B^2) +
\beta v^2_B \phi_B (1 - \phi_B^2/v_B^2),
\eeq
 which is still consistent with ``triviality'' (corresponding to an effective
potential given by the sum of a classical background and the zero point
energy of a free field) but allows for an explicit
scale-breaking term $\beta$. Setting
$\alpha=0$ one obtains a good description of the data in the ``extreme double
well'' limit ($r_o$ much more negative than $r_c$,
where $r_c$ corresponds to the onset of SSB)
where SSB is a semi-classical phenomenon and the
zero-point energy represents
a small perturbation. Then we start to increase $r_o$, at
fixed $\lambda_o$, toward the unknown value $r_c$ and look at
the quality of the fit with $(\alpha,\beta,v_B)$ as free parameters. The
minimum allowed value of $r_o$
 such that the quality
of the 2- parameter fit $(\alpha,\beta=0,v_B)$ is exactly the same as in the
more general 3- parameter case will define the ``massless'' case so that we can
fit the data for $J=J(\phi_B)$ with our Eq.(6).
Finally, Eq.(3) suggests that the vacuum field $v_B$, in lattice units,
 vanishes extremely fast for $\lambda_B \to 0$.
 Thus, to avoid that noise and signal become comparable,
 we cannot run the lattice simulation
at very small values of $\lambda_o$ but have to restrict to values
$\lambda_o \sim 1$
such that still
${{\lambda_o}\over{\pi^2}}<<1$ but $v_B$, in lattice units, is not too
small. Smaller values of $\lambda_o$ ($\sim$0.3-0.4),
however, should become accessible with the
largest lattices available today ($\sim100^4$) where one should safely
reach values $|J|a^3\sim$0.001 being still in agreement with Eq.(8).
 At $\lambda_o=1$
we have identified the massless regime at a
 value  $r_o=r_s$ where $r_sa^2 \sim -0.45$.
By using the accurate weak-coupling
relation between the bare mass and the euclidean cutoff \cite{CW}
\beq
 r_s=
-{{\lambda_B}\over{32\pi^2}}\Lambda^2
=-{{3\lambda_o}\over{16\pi^2}}\Lambda^2
\eeq
and using the relation $\Lambda={{\pi y_{Q}}\over{a}}$
(where $y_{Q}$ is expected to be O(1))
we obtain $y_{Q}\sim 1.55$. Also, the massless relation (3) predicts,
\beq
  (av_B)^{\rm 1-loop}=
{{\pi y_{L}}\over{\sqrt{3\lambda_o }}} \exp(-{{8\pi^2}\over{9\lambda_o}})
\eeq
and
\beq
Z^{\rm 1-loop}_{\phi}={{8\pi^2}\over{3\lambda_o}}.
\eeq
In Eq.(12) we have used $y_{L}$ rather than $y_{Q}$ since one
does not expect precisely the same coefficient to govern
the relation between euclidean cutoff and lattice spacing for both
quadratic and  logarithmic divergences (see below).
\par We have used an ALPHA-VAX to compute
with the Metropolis algorithm on a $10^4$ lattice.
We started by
comparing our results with those obtained by Huang et al.
\cite{huang} and  we found excellent
agreement (better than $1\%$) with their curves $J=J(\phi_B)$
in the range $|J|a^3\geq 0.05$. The other numerical results of \cite{huang},
however,
should be carefully reanalyzed due to two reasons
: 1) they have directly
 evaluated the 1- and
2- point Green's functions at $J=0$. As discussed above, the
computations at low $J$ have large errors due to the finite size of
the lattice producing unphysical violations of Eq.(8)
and invalidating all calculations at $J=0$ ;
2) there are uncontrolled errors from their
assumption
$Z_{\phi}=Z_h$ in the
evaluation of the shifted field propagator.
\par Our numerical values for $\phi_B(J)$ and the
results of the 2-parameter fits to the data by using Eq.(6)
 for $\lambda_o=$ 0.8,1.0 and 1.2 are shown in Table I together with the
one-loop prediction (13). The change  of $r_o$
with $\lambda_o$ is computed
 by using Eq.(11) and our result $r_sa^2\sim -0.45$ for $\lambda_o=1$.
After 60,000 iterations,
 which corresponds to the results of Table I, the average field
$\phi_B(J)$ is stable at the level of the first three significant digits
for all $J$-values. At large $J$, the stability is at the level of a few
$10^{-4}$ and finite size effects are negligible. This has been
checked with a few runs on a $16^4$ lattice and by comparing with Eq.(8). At
lower $J$, comparison with eq.(8) suggests that the third digit is affected by
finite size effects.
Many more details of the lattice calculation,
including a critical comparison
between Monte Carlo and the Langevin formulation of the lattice theory,
will be reported elsewhere \cite{ago}.
\par By fixing $Z_{\phi}$ to its one-loop value (13) and using Eq.(12) for
$av_B$
we can determine the
value of $y_{L}$ from the data at $\lambda_o=1$. We obtain
$y_L=2.067\pm 0.007$ with a ${{\chi^2}\over{d.o.f.}}={{5}\over{21}}$. Once we
know $y_L$ we can predict the value of $av_B=(av_B)^{\rm Th}$ at
$\lambda_o$=0.8 and 1.2 . The comparison with the 1-parameter fits in Table II
shows a remarkable agreement between one loop predictions and
Monte Carlo data. This result is not
a trivial test of perturbation theory but, rather, a non
perturbative test of ``triviality''. In fact, we have compared with the
leading-logarithmically ``improved'' version of Eq.(6)
\[
 J^{{\rm LL}}(\phi_B)={ {\lambda_o\phi^3_B}\over{1+{{9\lambda_o}
\over{8\pi^2}}
\ln{ { \pi x_{{\rm LL}} }\over{a|\phi_B|}} }}
\]
($x_{{\rm LL}}$ denoting an adjustable parameter). We find, for 21 degrees
of freedom,
 $(\chi^2)_{{\rm LL}}= 53,163,365$ for $\lambda_o=0.8,1.0,1.2$ respectively,
with an unacceptably low confidence level (less than 0.001 in the best
case). Also, the low $\chi^2$- values of the one-loop 1-parameter
fits suggest that, probably, we are still overestimating the
errors and the importance of the finite size effects.
\par In conclusion, the lattice computations nicely confirm the conjecture
\cite{consteve,zeit} that, as a consequence of ``triviality'', the one-loop
potential is exact. Indeed, there is a well defined region in the space of the
bare parameters $(r_o,\lambda_o)$, controlled by Eq.(11), where the effective
potential is described by its one-loop approximation to very high accuracy.
By increasing the lattice size, our analysis can be further extended towards
the physically relevant point ${{\lambda_o}\over{\pi^2}}\to 0^+$,
$r_oa^2\to 0^-$ which
corresponds, in the limit of quantum field theory, to
``dimensional transmutation'' \cite{CW} from the classically scale invariant
case.
The massless version
 of $\lambda\Phi^4$ theories, although ``trivial'' is not entirely
trivial providing, at the
same time, SSB and
 a meaningful continuum limit $\Lambda \to \infty$, $\lambda_B \to 0^+$ such
that the mass of the free shifted field in Eq.(3) is cutoff independent.
 As such it represents the real candidate to
generate SSB in the standard model from the pure scalar sector with $v_R$ in
Eqs.(4,5) related to the Fermi constant.
 The consequences of our results are substantial. As discussed
in \cite{consteve,zeit,con,iban,new}, and first pointed out in
\cite{huang},
the magnitude of the Higgs mass, beyond perturbation theory,
 does not represent a measure of the observable
interactions in the scalar sector of the theory.
\vskip 15 pt
\centerline{AKNOWLEDGEMENTS}
\par We thank K. Huang, A. Patrascioiu
 and P. M. Stevenson for very useful discussions and
collaboration.

\vfill
\eject
\widetext
\begin{table}
\caption
{ The values of $\phi_B(J)=\langle\Phi_B\rangle_J$ for the massless case
are reported as discussed in the text. At the various values of $\lambda_o$
and $r_o$
we also show the results of the 2-parameter
fit with Eq.(6) and the one loop prediction (13).}
\label{Table I}
\end{table}
\begin{tabular}{lccc}
{}~~~$Ja^3$ &~~~~ {$\lambda_o=0.8~r_oa^2=-0.36$}~~~ &~~~
 {$\lambda_o=1.0~r_oa^2=-0.45$}
{}~~~ & ~~~
{$\lambda_o=1.2~r_oa^2=-0.54$} \\
\tableline
-0.700 &$-1.0024\pm0.3\cdot10^{-3}$ & $-0.9401\pm 0.3\cdot10^{-3}$
&$-0.8935\pm0.3\cdot10^{-3}$ \\
\tableline
-0.600 &$-0.9540\pm0.3\cdot10^{-3}$ & $-0.8950\pm 0.3\cdot10^{-3}$
&$-0.8512\pm0.3\cdot10^{-3}$ \\
\tableline
-0.500 &$-0.8997\pm0.5\cdot10^{-3}$ & $-0.8444\pm 0.3\cdot10^{-3}$
&$-0.8037\pm0.3\cdot10^{-3}$ \\
\tableline
-0.400 &$-0.8371\pm0.5\cdot10^{-3}$ & $-0.7867\pm 0.5\cdot10^{-3}$
&$-0.7493\pm0.5\cdot10^{-3}$ \\
\tableline
-0.300 &$-0.763\pm1.0\cdot10^{-3}$ & $-0.718\pm 1.0\cdot10^{-3}$
&$-0.685\pm1.0\cdot10^{-3}$ \\
\tableline
-0.200 &$-0.670\pm1.0\cdot10^{-3}$ & $-0.631\pm 1.0\cdot10^{-3}$
&$-0.603\pm1.0\cdot10^{-3}$ \\
\tableline
-0.150 &$-0.611\pm1.5\cdot10^{-3}$ & $-0.576\pm 1.0\cdot10^{-3}$
&$-0.551\pm1.0\cdot10^{-3}$ \\
\tableline
-0.125&$-0.576\pm1.5\cdot10^{-3}$ & $-0.544\pm 1.5\cdot10^{-3}$
&$-0.521\pm1.0\cdot10^{-3}$ \\
\tableline
-0.100 &$-0.537\pm2.0\cdot10^{-3}$ & $-0.508\pm 2.0\cdot10^{-3}$
&$-0.486\pm2.4\cdot10^{-3}$ \\
\tableline
-0.075 &$-0.490\pm2.4\cdot10^{-3}$ & $-0.464\pm 2.0\cdot10^{-3}$
&$-0.446\pm2.4\cdot10^{-3}$ \\
\tableline
-0.050 &$-0.430\pm2.6\cdot10^{-3}$ & $-0.409\pm 3.0\cdot10^{-3}$
&$-0.393\pm2.4\cdot10^{-3}$ \\
\tableline
0.050 &$0.427\pm2.6\cdot10^{-3}$ & $0.406\pm 3.0\cdot10^{-3}$
&$0.391\pm2.4\cdot10^{-3}$ \\
\tableline
0.075 &$0.488\pm2.4\cdot10^{-3}$ & $0.462\pm 2.0\cdot10^{-3}$
&$0.444\pm2.4\cdot10^{-3}$ \\
\tableline
0.100&$0.535\pm2.0\cdot10^{-3}$ & $0.506\pm 2.0\cdot10^{-3}$
&$0.484\pm2.4\cdot10^{-3}$ \\
\tableline
0.125 &$0.575\pm1.5\cdot10^{-3}$ & $0.543\pm 1.5\cdot10^{-3}$
&$0.520\pm1.0\cdot10^{-3}$ \\
\tableline
0.150&$0.609\pm1.5\cdot10^{-3}$ & $0.575\pm 1.0\cdot10^{-3}$
&$0.550\pm1.0\cdot10^{-3}$ \\
\tableline
0.200 &$0.669\pm1.0\cdot10^{-3}$ & $0.631\pm 1.0\cdot10^{-3}$
&$0.602\pm1.0\cdot10^{-3}$ \\
\tableline
0.300 &$0.762\pm1.0\cdot10^{-3}$ &$0.717\pm 1.0\cdot10^{-3}$
&$0.684\pm1.0\cdot10^{-3}$ \\
\tableline
0.400 &$0.8367\pm0.5\cdot10^{-3}$ & $0.7863\pm 0.5\cdot10^{-3}$
&$0.7488\pm0.5\cdot10^{-3}$ \\
\tableline
0.500 &$0.8993\pm0.5\cdot10^{-3}$ &$0.8445\pm 0.3\cdot10^{-3}$
&$0.8035\pm0.3\cdot10^{-3}$ \\
\tableline
0.600 &$0.9540\pm0.3\cdot10^{-3}$ &$0.8951\pm 0.3\cdot10^{-3}$
&$0.8513\pm0.3\cdot10^{-3}$ \\
\tableline
0.700 &$1.0024\pm0.3\cdot10^{-3}$ &$0.9402\pm 0.3\cdot10^{-3}$
&$0.8937\pm0.3\cdot10^{-3}$\\
\tableline
{}~~~~~~&$Z_{\phi}=33.3\pm 0.6$ &$Z_{\phi}=26.0\pm0.4$
&$Z_{\phi}=21.3\pm0.3$\\
\tableline
{}~~~~~~&$av_B=(5.9\pm 2.0)10^{-5}$ &$av_B=(6.9\pm 1.4)10^{-4}$
&$av_B=(3.1\pm0.4)10^{-3}$\\
\tableline
{}~~~~~~&$Z^{\rm 1-loop}_{\phi}=32.9$ &$Z^{\rm 1-loop}_{\phi}=26.3$
&$Z^{\rm 1-loop}_{\phi}=21.9$\\
\end{tabular}

\vskip 50pt

\begin{table}
\caption{By using Eq.(6), we show the results of the 1-parameter fits
for $av_B$ at $\lambda_o=$0.8 and 1.2 when
$Z_{\phi}$ is constrained to its one-loop value in Eq.(13). We also show the
predictions from Eq.(12), $(av_B)^{\rm Th}$, for $y_L=2.067\pm0.007$ as
determined from the
fit to the data at $\lambda_o=$1.}
\label{Table II}
\end{table}
\begin{tabular}{cc}
\tableline
{}~~~~~~~~~~~~~~~~{$\lambda_o=0.8~~r_oa^2=-0.36$}~~~~~~~~~~~~~~~~~~~~~ &
{$\lambda_o=1.2~~r_oa^2=-0.54$} \\
\tableline
$Z_{\phi}=32.90=fixed$ &$Z_{\phi}=21.93=fixed$\\
\tableline
$av_B=(7.30\pm0.03)10^{-5}$ &$av_B=(2.28\pm0.01)10^{-3}$\\
\tableline
${{\chi^2}\over{d.o.f}}={{5}\over{21}}$&${{\chi^2}\over{d.o.f}}=
{{13}\over{21}}$\\
\tableline
$(av_B)^{\rm Th}=(7.24\pm0.03)10^{-5}$ &
$(av_B)^{\rm Th}=(2.29\pm0.01)10^{-3}$
\end{tabular}
\end{document}